\documentclass[twocolumn,prb,showpacs,superscriptaddress,groupedaddress]{revtex4}
\usepackage{epsfig,color}
\usepackage{amssymb}
\usepackage{amsmath}
\usepackage{dcolumn}
\usepackage{graphicx}

\usepackage[pdfborder={0 0 0},colorlinks=true,linkcolor=blue,citecolor=blue]{hyperref}

\def\beq{\begin{equation}}
\def\eeq{\end{equation}}
\def\bea{\begin{eqnarray}}
\def\eea{\end{eqnarray}}

\def\q{\mathbf{q}}
\def\k{\mathbf{k}}
\def\p{\mathbf{p}}

\def\Q{\mathbf{Q}}

\newcommand{\eps}{\varepsilon}

\newcommand{\su}{\uparrow}
\newcommand{\sd}{\downarrow}

\newcommand{\sign}{\mathrm{sgn}}

\newcommand{\ii}{{\mathrm{i}}}

\newcommand{\nn}{\nonumber}

\newcommand{\GDLL}[1]{\textcolor{red}{#1}}
\newcommand{\DLS}[1]{\textcolor[rgb]{0,0.9,0}{#1}}
\newcommand{\DLL}[1]{\textcolor[rgb]{0.9,0.7,0}{#1}}

\begin{document}

\title{Spin resonance peak in Fe-based superconductors with unequal gaps}
\author{M.M.~Korshunov}
\email{mkor@iph.krasn.ru}
\affiliation{Kirensky Institute of Physics, Federal Research Center KSC SB RAS, 660036 Krasnoyarsk, Russia}
\affiliation{Siberian Federal University, Svobodny Prospect 79, 660041 Krasnoyarsk, Russia}
\author{V.A.~Shestakov}
\affiliation{Siberian Federal University, Svobodny Prospect 79, 660041 Krasnoyarsk, Russia}
\author{Yu.N.~Togushova}
\affiliation{Siberian Federal University, Svobodny Prospect 79, 660041 Krasnoyarsk, Russia}

\date{\today}

\begin{abstract}
We study the spin resonance in superconducting state of iron-based materials within multiband models with two unequal gaps, $\Delta_L$ and $\Delta_S$, on different Fermi surface pockets. We show that due to the indirect nature of the gap entering the spin susceptibility at the nesting wave vector $\Q$ the total gap $\tilde\Delta$ in the bare susceptibility is determined by the sum of gaps on two different Fermi surface sheets connected by $\Q$. For the Fermi surface geometry characteristic to the most of iron pnictides and chalcogenides, the indirect gap is either $\tilde\Delta = \Delta_L + \Delta_S$ or $\tilde\Delta = 2\Delta_L$. In the $s_{++}$ state, spin excitations below $\tilde\Delta$ are absent unless additional scattering mechanisms are assumed. The spin resonance appears in the $s_\pm$ superconducting state at frequency $\omega_R \leq \tilde\Delta$. Comparison with available inelastic neutron scattering data confirms that what is seen is the true spin resonance and not a peak inherent to the $s_{++}$ state.
\end{abstract}

\maketitle

\section{Introduction}

Fe-based superconductors (FeBS) represent a non-cuprate class of high-$T_c$ systems with the unconventional superconducting state. The origin of the latter is still debated. In general, FeBS can be divided into the two subclasses, pnictides and chalcogenides~\cite{Reviews}, with the square lattice of iron as the basic element, though with orthorhombic distortions in lightly doped materials. 
Iron is surrounded by As or P situated in the tetrahedral positions within the first subclass and by Se, Te, or S within the second subclass.

Fermi surface (FS) is formed by Fe $d$-orbitals and excluding the cases of extreme hole and electron dopings it consists of two hole sheets around the $\Gamma=(0,0)$ point and two electron sheets around the $(\pi,0)$ and $(0,\pi)$ points in the two-dimensional Brillouin zone (BZ) corresponding to one Fe per unit cell (the so-called 1-Fe BZ)~\cite{ROPPreview2011}. In the 2-Fe BZ, electron pockets are centered at the $M=(\pi,\pi)$ point. Nesting between these two groups of sheets leads to the enhanced antiferromagnetic fluctuations with the maximal scattering near the wave vector $\Q$ equal to $(\pi,0)$ or $(0,\pi)$ in the 1-Fe BZ or to $(\pi,\pi)$ in the 2-Fe BZ.


Different mechanisms of Cooper pairs formation result in the distinct superconducting gap symmetry and structure~\cite{ROPPreview2011}. In particular, the RPA-SF (random-phase approximation spin fluctuation) approach gives the extended $s$-wave gap that changes sign between hole and electron FS sheets ($s_{\pm}$ state) as the main instability for the wide range of doping concentrations~\cite{Mazin_etal_splusminus,Graser2009,Kuroki2008,MaitiPRB,KorshunovUFN}. On the other hand, orbital fluctuations promote the order parameter to have the sign-preserving $s_{++}$ symmetry~\cite{Kontani}. Thus, probing the gap structure can help in elucidating the underlying mechanism. In this respect, inelastic neutron scattering (INS) is a useful tool since the measured dynamical spin susceptibility $\chi(\q,\omega)$ in the superconducting state carries information about the gap structure. There are been many reports of a well-defined peak in neutron spectra in 1111, 122, and 11 systems appearing only for $T < T_c$ at or around $\q = \Q$~\cite{ChristiansonBKFA,Inosov2010,Argyriou2010,Lumsden2011,Dai2015}. The common explanation is that the peak is the spin resonance appearing due to the $s_\pm$ state. Indeed, since $\Q$ connects Fermi sheets with different signs of $s_\pm$ gaps, the resonance condition for the interband susceptibility is fulfilled and the spin resonance peak is formed at a frequency $\omega_R$ below $\approx 2\Delta$ with $\Delta$ being the gap size~\cite{Korshunov2008,Maier,Maier2}.


Such simple explanation was indirectly questioned by the angle-resolved photoemission spectroscopy (ARPES) results and recent measurements of gaps via Andreev spectroscopy. Latter clearly shows that there are at least two distinct gaps present in 11, 122, and 1111 systems~\cite{Daghero2009,Tortello2010,Ponomarev2013,Abdel-Hafiez2014,Kuzmichev2016} and even three gaps in LiFeAs~\cite{Kuzmichev2012,Kuzmichev2013}. Larger gap ($\Delta_L$) is about 9meV and the smaller gap ($\Delta_S$) is about 4meV in BaCo122 materials. From ARPES we know that electron FS sheets and the inner hole sheet are subject to opening the lager gap while the smaller gap is located at the outer hole FS~\cite{Ding2008,Evtushinsky2009}. The very existence of the smaller gap rise the question -- what would be the spin resonance frequency in the system with two distinct gaps? Naive expectation is that the frequency shifts to the lower gap scale and $\omega_R < 2 \Delta_S$. Then the observed peak in INS in BaCo122 system at frequency $\omega_{INS} \sim 9.5$meV~\cite{Inosov2010} can not be the spin resonance since it is greater than $2\Delta_S \sim 8$meV~\cite{Tortello2010}. Thus the peak could be coming from the $s_{++}$ state~\cite{Onari2010,Onari2011}, where it forms at frequencies \textit{above} $2\Delta$ due to the redistribution of the spectral weight upon entering the superconducting state and a special form of scattering in the normal state. Here we study this question in details and show that the naive expectation is wrong and that the true minimal energy scale is $\omega_R \leq \Delta_L+\Delta_S$. Latter is consistent with the maximal frequency of the observed peak in INS in BaCo122 and confirms that it is the true spin resonance evidencing the $s_\pm$ gap symmetry. The maximal energy scale is $\omega_R \leq 2\Delta_L$. Whether the minimal or maximal energy scale will be realized depends on the relation between the exact band structure of a particular material and the wave vector of the spin resonance $\Q$.

\section{Models and approach}

To describe spin response in normal and superconducting states of FeBS, we use random phase approximation (RPA) with the local Coulomb interactions (Hubbard and Hund's exchange). In the multiband system, transverse dynamical spin susceptibility $\hat\chi_{+-}(\q,\omega)$ is the matrix in orbital (or band) indices. It can be obtained in the RPA from the bare electron-hole matrix bubble $\hat\chi_{(0)+-}(\q,\omega)$ by summing up a series of ladder diagrams:
\begin{eqnarray}
\hat\chi_{+-}(\q,\omega) = \left[\hat{I} - \hat{U}_s \hat\chi_{(0)+-}(\q,\omega)\right]^{-1} \hat\chi_{(0)+-}(\q,\omega),
\label{eq:chi_s_sol}
\end{eqnarray}
where $\q$ is the momentum, $\omega$ is the frequency, $\hat{U}_s$ and $\hat{I}$ are interaction and unit matrices in orbital (or band) space. Exact form of $\hat{U}_s$ and bare susceptibility $\hat\chi_{(0)+-}(\q,\omega)$ depends on the underlying model. Later we use two types of tight-binding models for the two-dimensional iron layer.

First we study the four-band model of Ref.~\onlinecite{Korshunov2008} with the following single-electron Hamiltonian
\begin{eqnarray}
 H_0 = - \sum\limits_{\k,\alpha ,\sigma } {{\epsilon^i} n_{\k i \sigma}} - \sum\limits_{\k, i, \sigma}  t_{\k}^{i} d_{\k i \sigma}^\dag d_{\k i \sigma}, \label{eq:H04band}
\end{eqnarray}
where $d_{\k i \sigma}$ is the annihilation operator of the $d$-electron with momentum $\k$, spin $\sigma$, and band index $i = \left\{ \alpha_1, \alpha_2, \beta_1, \beta_2 \right\}$, $\epsilon^i$ are the on-site single-electron energies,
$t_{\k}^{\alpha_{1,2}}$ is the electronic dispersion that yields hole pockets centered around the $\Gamma$ point, and $t_{\k}^{\beta_{1,2}}$ is the dispersion that results in the electron pockets around the $M$ point of the 2-FeBZ. Parameters are the same as in Ref.~\onlinecite{Korshunov2008}.
In the superconducting state we assume either the $s_{++}$ state with $\Delta_{\k i} = \Delta_{i}$ or the $s_\pm$ state with $\Delta_{\k i} = \Delta_{i} \left( \cos k_x + \cos k_y \right) / 2$.

Physical spin susceptibility $\chi(\q,\omega) = \sum_{i,j}\chi^{i,j}(\q,\omega)$ obtained by calculating matrix elements $\chi^{i,j}(\q,\omega)$ via equation~(\ref{eq:chi_s_sol}) with the interaction matrix $U_s^{i,j} = \tilde{U} \delta_{i,j} + \tilde{J}/2 (1-\delta_{i,j})$ and with the bare spin susceptibility $\chi_{(0)+-}^{ij}(\q,\omega)$ in the superconducting state (see Ref.~\onlinecite{Korshunov2008} for details).
%
We assume here the effective Hubbard interaction parameters to be $\tilde{J}=0.2 \tilde{U}$ and $\tilde{U} \sim t^{\beta_1}_1$ in order to stay in the paramagnetic phase~\cite{Korshunov2008}.

The model described above is simple enough to gain qualitative description of the spin response of superconductor with unequal gaps. But it lack for the orbital content of the bands that is important for the detailed structure of the susceptibility. That is why we also present results for the tight-binding model from Ref.~\onlinecite{Graser2009} based on the fit to the DFT band structure for LaFeAsO~\cite{Cao2008}. The model includes all five iron $d$-orbitals ($d_{xz}$, $d_{yz}$, $d_{xy}$, $d_{x^2-y^2}$, $d_{3z^2-r^2}$) enumerated by index $l$
and is given by
\begin{equation}
 H_0 = \sum_{\k \sigma} \sum_{l l'} \left[ t_{l l'}(\k) + \epsilon_{l} \delta_{l l'} \right] d_{l \k \sigma}^\dagger d_{l' \k \sigma},
 \label{eq:H0}
\end{equation}
where $d_{l \k \sigma}^\dagger$ is the annihilation operator of a particle with momentum $\k$, spin $\sigma$, and orbital index $l$. Later we use numerical values of hopping matrix elements $t_{l l'}(\k)$ and one-electron energies $\epsilon_{l}$ from Ref.~\onlinecite{Graser2009}. This model for the undoped and moderately electron doped materials gives FS composed of two hole pockets, $\alpha_1$ and $\alpha_2$, around the $(0,0)$ point and two electron pockets, $\beta_1$ and $\beta_2$, centered around $(\pi,0)$ and $(0,\pi)$ points of the 1-Fe BZ. 
Similar model for iron pnictides was proposed in Ref.~\onlinecite{Kuroki2008}.

The general two-particle on-site interaction would be represented by the Hamiltonian~\cite{Castallani1978,Oles1983,Kuroki2008,Graser2009}:
\bea
H_{int} &=& U \sum_{f, m} n_{f m \su} n_{f m \sd} + U' \sum_{f, m < l} n_{f l} n_{f m} \nn\\
  && + J \sum_{f, m < l} \sum_{\sigma,\sigma'} d_{f l \sigma}^\dag d_{f m \sigma'}^\dag d_{f l \sigma'} d_{f m \sigma} \nn\\
  && + J' \sum_{f, m \neq l} d_{f l \su}^\dag d_{f l \sd}^\dag d_{f m \sd} d_{f m \su}.
\label{eq:Hint}
\eea
where $n_{f m} = n_{f m \su} + n_{f m \sd}$, $n_{f m \sigma} = d_{f m \sigma}^\dag d_{f m \sigma}$ is the number of particles operator at the site $f$, $U$ and $U'$ are the intra- and interorbital Hubbard repulsion, $J$ is the Hund's exchange, and $J'$ is the so-called pair hopping. We choose the following values for the interaction parameters: $U=1.4$eV, $J=0$, and make use of the spin-rotational invariance constraint $U'=U-2J$ and $J'=J$.


Green functions are diagonal in the band basis but not in the orbital basis. Let us introduce creation and annihilation operators $b_{\k \mu \sigma}^\dag$ and $b_{\k \mu \sigma}$ of electrons with band index $\mu$, in terms of which Green functions are diagonal, $G_{\mu \sigma}(\k,\ii\Omega) = 1 / \left( \ii\Omega - \eps_{\k\mu\sigma} \right)$. Transformation from the orbital to the band basis is done via the matrix elements $\varphi^{\mu}_{\k m}$,
$d_{\k m \sigma} = \sum\limits_{\mu} \varphi^{\mu}_{\k m} b_{\k \mu \sigma}$, and
for the transverse component of the bare spin susceptibility~\cite{KorshunovUFN} we have
%
\bea \label{eq:chipmmu}
 &&\chi^{ll',mm'}_{(0)+-}(\q,\ii\Omega) = -T \sum_{\p,\omega_n, \mu,\nu} \left[ \varphi^{\mu}_{\p m} {\varphi^*}^{\mu}_{\p l} G_{\mu \su}(p,\ii\omega_n) \right.\nn\\
 &&\times G_{\nu \sd}(\p+\q,\ii\Omega+\ii\omega_n) \varphi^{\nu}_{\p+\q l'} {\varphi^*}^{\nu}_{\p+\q m'} \nn\\
 &&- {\varphi^*}^{\mu}_{\p l} {\varphi^*}^{\mu}_{-\p m'} F^\dag_{\mu \su}(\p,-\ii\omega_n) \nn\\
 &&\left. \times F_{\nu \sd}(\p+\q,\ii\Omega+\ii\omega_n) \varphi^{\nu}_{\p+\q l'} \varphi^{\nu}_{-\p-\q m} \right],
\eea
where $\Omega$ and $\omega_n$ are Matsubara frequencies, $G$ and $F$ are the normal and anomalous (superconducting) Green's functions, respectively.
Components of the physical spin susceptibility $\chi_{+-}(\q,\ii\Omega) = \frac{1}{2} \sum_{l,m} \chi^{ll,mm}_{+-}(\q,\ii\Omega)$ are calculated using Eq.~(\ref{eq:chi_s_sol}) with the interaction matrix $U_s$ from Ref.~\onlinecite{Graser2009}.

Since calculation of the Cooper pairing instability is not a topic of the present study, here we assume that the superconductivity is coming from some other theory and study either the $s_{++}$ state with $\Delta_{\k\mu} = \Delta_{\mu}$ or the $s_\pm$ state with $\Delta_{\k\mu} = \Delta_{\mu} \cos k_x \cos k_y$, where $\mu$ is the band index.

\section{Results}

Here we present results for susceptibilities at the wave vector $\q=\Q$ as functions of frequency $\omega$ obtained via analytical continuation from Matsubara frequencies ($\ii\Omega \to \omega + \ii\delta$ with $\delta \to 0+$).

\begin{figure}[ht]
\begin{center}
\includegraphics[width=\linewidth]{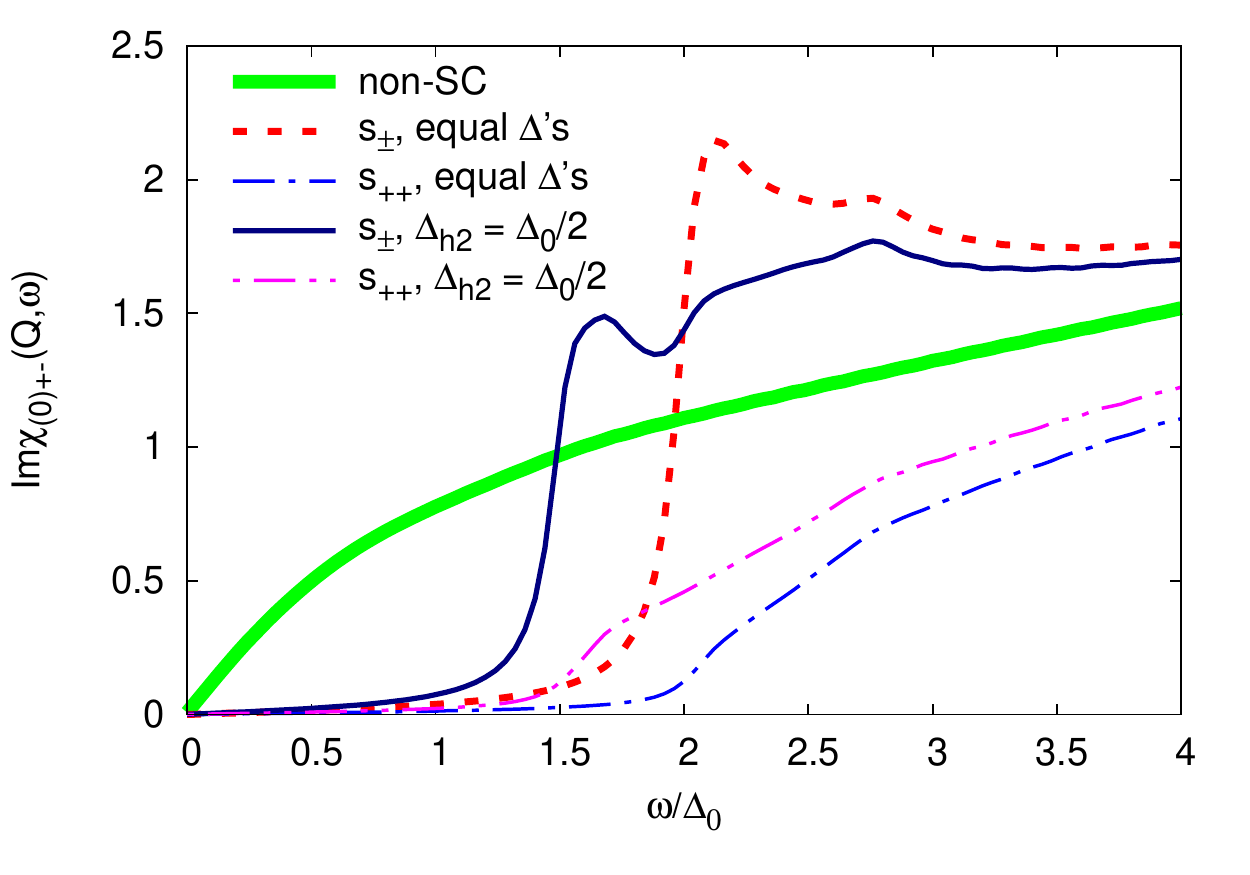}
\includegraphics[width=\linewidth]{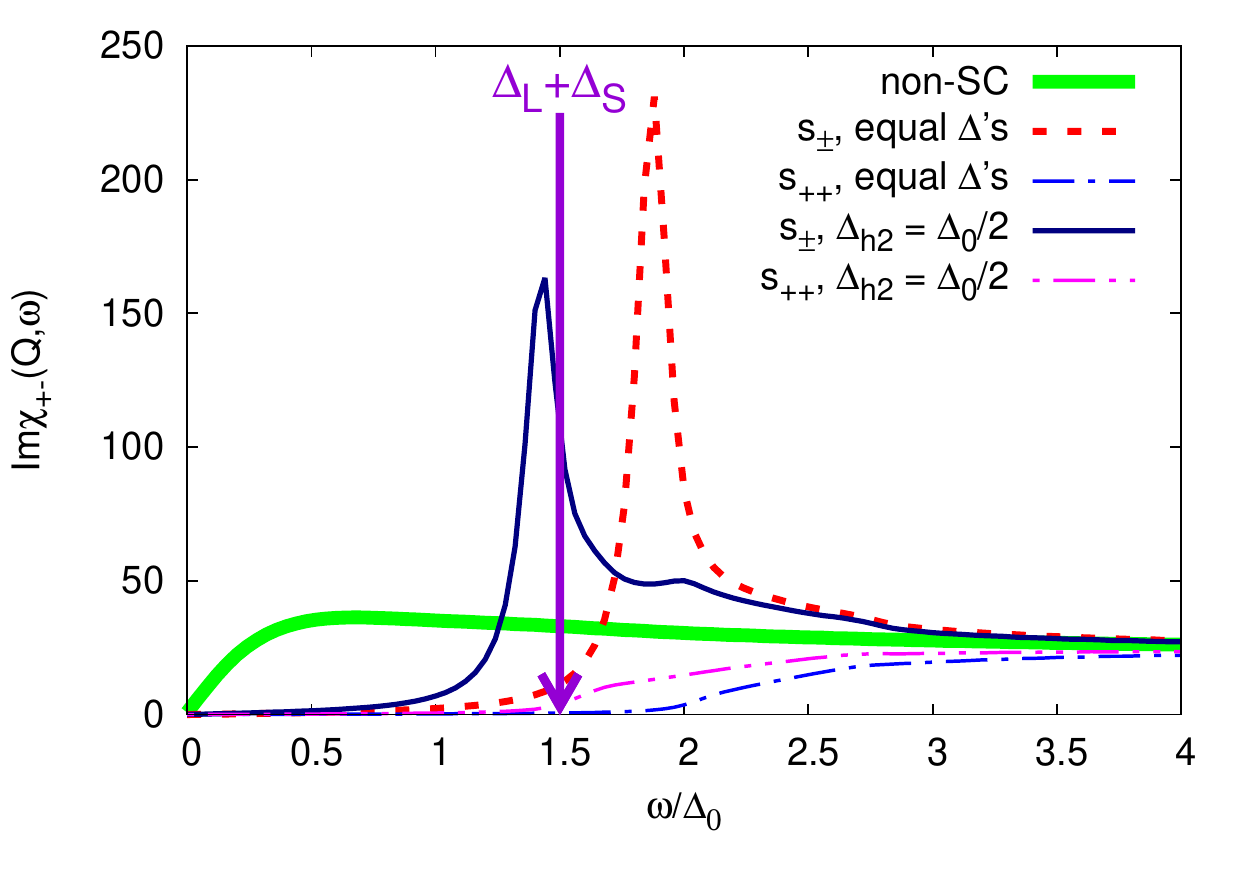}
\caption{(Color online) Calculated $\mathrm{Im}\chi_{(0)+-}(\Q,\omega)$ (top) and $\mathrm{Im}\chi_{+-}(\Q,\omega)$ with $\Q=(\pi,\pi)$ in the 2-Fe BZ for the four-band model in the normal, $s_{++}$ and $s_\pm$ superconducting states. Two cases of superconducting states are shown: equal $\Delta$'s with $\Delta_{e1,2} = \Delta_{h1,2} = \Delta_0$, and unequal gaps with $\Delta_{e1,2} = \Delta_{h1} = \Delta_0$ and $\Delta_{h2} = \Delta_0/2$.
\label{fig:ImChi4band}}
\end{center}
\end{figure}

Imaginary part of bare and RPA spin susceptibilities in the four-band model~(\ref{eq:H04band}) are shown in Fig.~\ref{fig:ImChi4band}. First, we discuss result for equal gaps on electron ($e1$, $e2$) and hole ($h1$, $h2$) FSs, $\Delta_{e1,2} = \Delta_{h1,2} = \Delta_0$.
Since $\chi_{(0)+-}(\q,\omega)$ describes particle-hole excitations and in the superconducting state all excitations are gapped below approximately $2\Delta_0$ (at $T=0$), then $\mathrm{Im}\chi_{(0)+-}(\q,\omega)$ becomes finite only after that frequency. For the $s_{++}$ state, there is a gradual increase of the spin response for $\omega > 2\Delta_0$. For the $s_\pm$ state, $\Q$ connects FSs with different signs of gaps, $\sign \Delta_{\k i} = - \sign \Delta_{\k+\Q j}$, and within RPA~(\ref{eq:chi_s_sol}) this results in the spin resonance peak -- divergence of $\mathrm{Im}\chi_{+-}(\Q,\omega)$ at a frequency $\omega_R < \omega_c$, see Fig.~\ref{fig:ImChi4band}, bottom panel.

Now let's consider the case of unequal gaps with a small gap scale on outer hole FS,  $\Delta_{h2} = \Delta_0/2$, and a larger gap scale on all other FSs, $\Delta_{e1,2} = \Delta_{h1} = \Delta_0$. As seen from Fig.~\ref{fig:ImChi4band}, top panel, for the $s_\pm$ state the discontinuous jump and, thus, $\omega_c$, moved to lower frequencies. This new energy scale clearly tracked down in the $s_{++}$ state as the starting point of the susceptibility gradual increase. It is equal to $\Delta_L + \Delta_S = 3/2 \Delta_0$, where $\Delta_L$ and $\Delta_S$ being the larger and smaller gap scales. Consequently, the spin resonance peak in $s_\pm$ moved to lower frequencies, $\omega_R < \Delta_L + \Delta_S$, see Fig.~\ref{fig:ImChi4band}, bottom panel. Additional feature is the hump around the $2\Delta_L = 2\Delta_0$ energy scale. Note that the susceptibility in the $s_{++}$ state haven't changed much compared to the equal gaps case.

\begin{figure}[ht]
\begin{center}
\includegraphics[width=\linewidth]{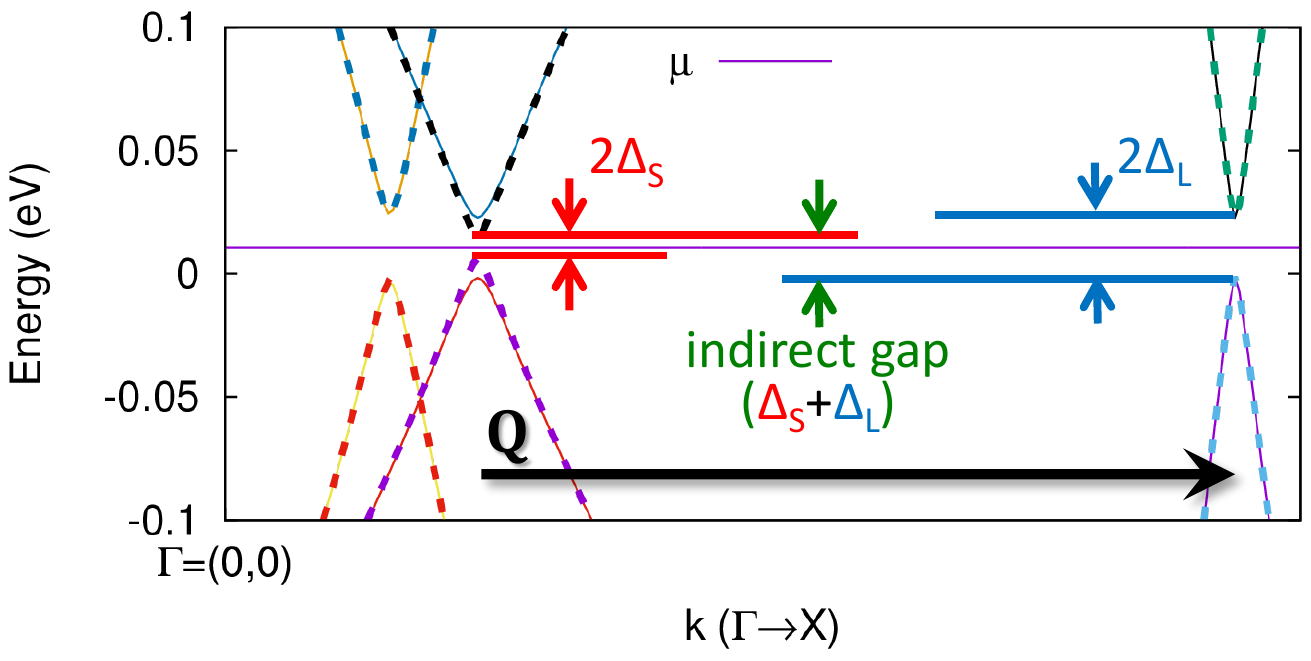}
\caption{(Color online) Energy spectrum of the five-orbital model near the Fermi level $\eps_F = \mu$ in the superconducting state, $E_{\k \nu} = \pm\sqrt{\eps_{\k \nu}^2 + \Delta_{\k \nu}^2}$, as a function of momentum $\k$ along the $\Gamma-X$ direction, i.e. $(0,0)-(\pi,0)$. Scattering wave vector $\Q$ entering the spin susceptibility is also shown.
\label{fig:5orbDelta}}
\end{center}
\end{figure}

\begin{figure}[ht]
\begin{center}
\includegraphics[width=\linewidth]{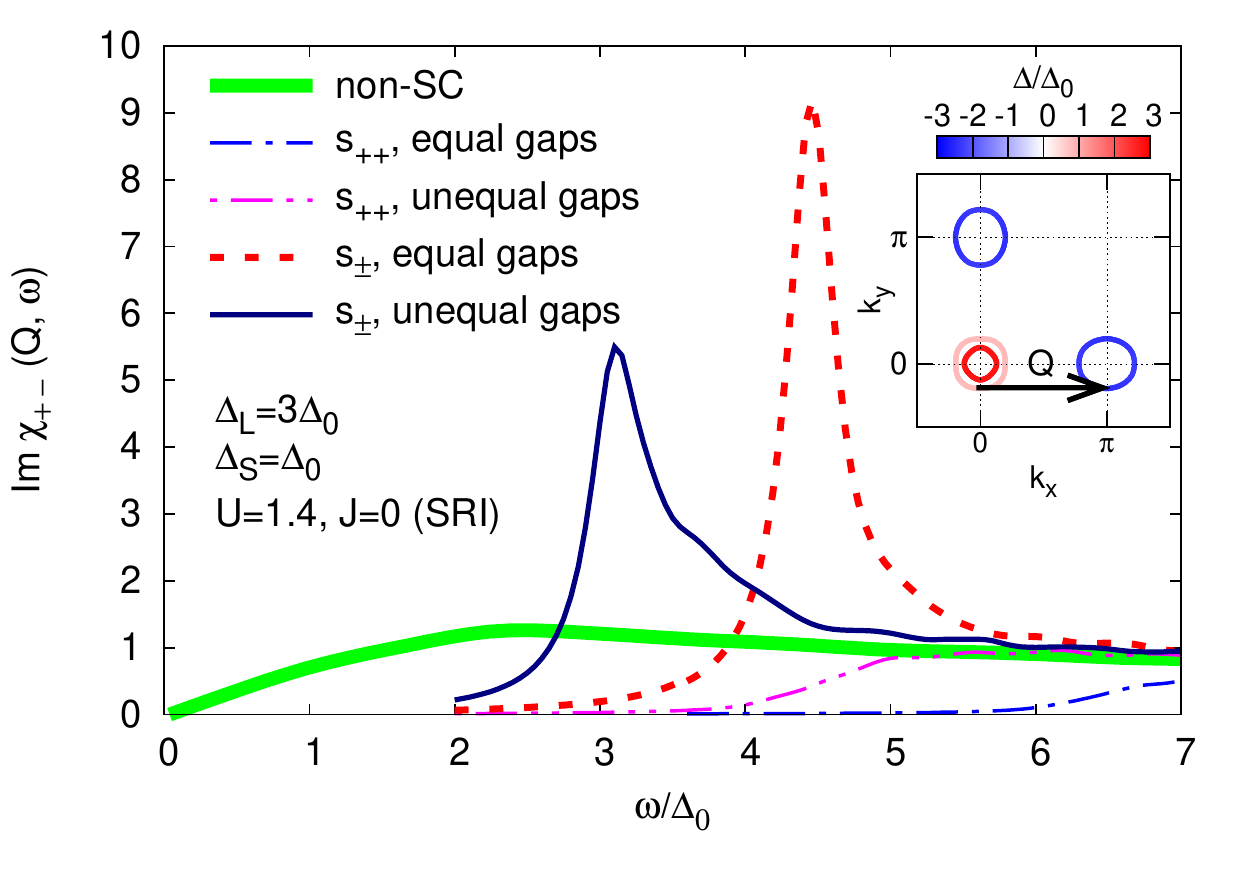}
\caption{(Color online) Calculated $\mathrm{Im}\chi_{+-}(\Q,\omega)$ with $\Q=(\pi,0)$ in the 1-Fe BZ for the five-orbital model in the normal, $s_{++}$ and $s_\pm$ superconducting states. Two cases of superconducting states are shown: equal gaps with $\Delta_{\alpha_{1,2}} = \Delta_{\beta_{1,2}} = \Delta_L$, and unequal gaps with $\Delta_{\alpha_{1,2}} = \Delta_{\beta_{1}} = \Delta_L$ and $\Delta_{\beta_2} = \Delta_S$, where $\Delta_S = \Delta_L / 3$. Latter case is shown in the inset, where gaps at the FS are plotted together with the wave vector $\Q$.
\label{fig:5orbImChi}}
\end{center}
\end{figure}

To demonstrate where the new energy scale is coming from we turn our attention to the five-orbital model~(\ref{eq:H0}). Its energy spectrum near the Fermi level in the superconducting state, $E_{\k \nu} = \pm\sqrt{\eps_{\k \nu}^2 + \Delta_{\k \nu}^2}$, is shown in Fig.~\ref{fig:5orbDelta}. We consider here the case of unequal gaps with the smaller gap $\Delta_{\beta_2} = \Delta_S$ on the outer hole FS and larger gaps $\Delta_{\alpha_{1,2}} = \Delta_{\beta_{1}} = \Delta_L$ on inner hole and electron FSs. To be consistent with the experimental data, we choose $\Delta_S = \Delta_0 = \Delta_L / 3$, see the inset in Fig.~\ref{fig:5orbImChi}. Naturally, the two energy scales, $2\Delta_S$ and $2\Delta_L$, appear in the energy spectrum $E_{\k \nu}$ and they are connected with hole $\alpha_2$ and electron $\beta_{1,2}$ bands, respectively. On the other hand, the susceptibility $\chi_{(0)+-}(\Q,\omega)$ contains scattering \textit{between} hole and electron bands with the wave vector $\Q$. The energy gap that have to be overcome to excite electron-hole pair is the indirect gap with the scale $\tilde\Delta = \Delta_L + \Delta_S$. That is why spin excitations in the $s_{++}$ state start with the frequency proportional to the indirect gap $\tilde\Delta = 4\Delta_0$, see Fig.~\ref{fig:5orbImChi}. The same is true for the discontinuous jump in $\mathrm{Im}\chi_{(0)}$ for the $s_\pm$ state -- it shifts to frequency $\approx \tilde\Delta$. This, together with the corresponding $\log$ singularity in $\mathrm{Re}\chi_{(0)}$, produce the spin resonance peak in RPA at frequency $\omega_R \leq \tilde\Delta$. Such shift of resonance peak to lower frequencies compared to the equal gaps situation is seen in Fig.~\ref{fig:5orbImChi}, where the spin response $\mathrm{Im}\chi_{+-}(\Q,\omega)$ for the cases of equal and distinct gaps is shown.

The changes in the band structure and/or doping level can result in the change of the indirect gap. In particular, since for the hole doping hole FSs become larger the wave vector $\Q$ may connect states on the electron FS and on the \textit{inner} hole FS. Gaps on both these FSs are determined by $\Delta_L$ and thus the indirect gap will be equal to $\tilde\Delta = 2\Delta_L$. This sets up a maximal energy scale for the spin resonance, i.e. $\omega_R \leq 2\Delta_L$.

Thus we conclude that depending on the relation between the wave vector $\Q$ and the exact FS geometry, the indirect gap in most FeBS can be either $\tilde\Delta = \Delta_L + \Delta_S$ or $\tilde\Delta = 2\Delta_L$. The peak in the dynamical spin susceptibility at the wave vector $\Q$ will be the true spin resonance if it appears below the indirect gap scale, $\omega_R \leq \tilde\Delta$.

Now we can compare energy scales extracted from ARPES, Andreev spectroscopy, and inelastic neutron scattering. Latter gives peak frequency $\omega_{INS} \approx 9.5$meV in BaFe$_{1.85}$Co$_{0.15}$As$_2$ with $T_c = 25$K~\cite{Inosov2010}. For the same system, gap sizes extracted from ARPES are $\Delta_L \approx 6.7$meV and $\Delta_S \approx 4.5$meV~\cite{Terashima2009}, and for a similar system with $T_c = 25.5$K, $\Delta_L \approx 6.6$meV and $\Delta_S \approx 5$meV~\cite{Kawahara2010}. Gap sizes extracted from Andreev spectroscopy are $\Delta_L \approx 9$meV and $\Delta_S \approx 4$meV in BaFe$_{1.8}$Co$_{0.2}$As$_2$ with $T_c = 24.5$K~\cite{Tortello2010}. Evidently, $\omega_{INS} < \Delta_L + \Delta_S$ and we can safely state that the peak in INS is the spin resonance.

For the hole doped systems, peak frequency in INS is about $14$meV in Ba$_{0.6}$K$_{0.4}$Fe$_2$As$_2$ with $T_c = 38$K~\cite{ChristiansonBKFA}. There is a slight discrepancy between gap sizes extracted from ARPES and Andreev spectra. Former gives $\Delta_L \approx 12$meV and $\Delta_S \approx 6$meV in the same material with $T_c = 37$K~\cite{Ding2008}, thus $\omega_{INS} < \Delta_L + \Delta_S$. Gap sizes from Andereev spectroscopy are $\Delta_L \approx 8$meV and $\Delta_S \approx 2$meV in Ba$_{0.65}$K$_{0.35}$Fe$_2$As$_2$ with lower $T_c = 34$K~\cite{Abdel-Hafiez2014}. In this case, $\omega_{INS} > \Delta_L + \Delta_S$ but $\omega_{INS} < 2\Delta_L$ and we still can assume that the peak in INS is the spin resonance. However, in such a case definitive conclusion can be given only by the calculation of spin response for the particular experimental band structure.
%
%
For more extensive review of available experimental data on $\omega_{INS}$ and gap scales, see the Supplemental Material~\cite{EPAPS}.

On the separate note, we would like to mention that the appearance of a hump structure in the superconducting state at frequencies larger than the main peak frequency (the so-called double resonance feature) may be related to the $2\Delta_L$ energy scale, see Fig.~\ref{fig:ImChi4band}. Such hump structure was observed in NaFe$_{0.985}$Co$_{0.015}$As~\cite{Zhang.PhysRevLett.111.207002.pdf,Zhang.PhysRevB.90.140502.pdf} and FeTe$_{0.5}$Se$_{0.5}$~\cite{Mook.PhysRevLett.104.187002.pdf}. Somehow similar structure was found in polarized inelastic neutron studies of BaFe$_{1.9}$Ni$_{0.1}$As$_2$~\cite{Lipscombe.PhysRevB.82.064515.pdf} and Ba(Fe$_{0.94}$Co$_{0.06}$)$_2$As$_2$~\cite{Steffens.PhysRevLett.110.137001.pdf}, but its origin may be related to the spin-orbit coupling~\cite{Korshunov2013} rather than the simple $2\Delta_L$ energy scale. Another explanation of the double resonance feature is related to the pre-existing magnon mode, i.e. the dispersive low-energy peak in underdoped materials is associated with the spin excitations of the magnetic order with the intensity enhanced below $T_c$ due to the suppression of the damping~\cite{Wang2016}.

\section{Conclusion}

We analysed the spin response of FeBS with two different superconducting gap scales,  $\Delta_L > \Delta_S$. Spin resonance appears in the $s_\pm$ state below the indirect gap scale $\tilde\Delta$ that is determined by the sum of gaps on two different Fermi surface sheets connected by the scattering wave vector $\Q$. In the $s_{++}$ state, spin excitations are absent below $\tilde\Delta$ unless additional scattering mechanisms are assumed~\cite{Kontani}. For the Fermi surface geometry characteristic to the most of FeBS materials, the indirect gap is either $\tilde\Delta = \Delta_L + \Delta_S$ or $\tilde\Delta = 2\Delta_L$. This gives the simple criterion to determine whether the experimentally observed peak in inelastic neutron scattering is the true spin resonance -- if the peak frequency $\omega_R$ is less than the indirect gap $\tilde\Delta$, then it is the spin resonance and, consequently, the superconducting state has the $s_\pm$ gap structure.

Comparison of energy scales extracted from INS, Andreev spectroscopy, ARPES and other techniques allowing to determine superconducting gaps, for most materials gives confidence that the observed feature in INS is the spin resonance peak. However, sometimes it is not always clear experimentally which gaps are connected by the wave vector $\Q$. Even without knowing this exactly, one can draw some conclusions. For example, if one of the gaps is $\Delta_L$, then there are three cases possible: (1) $\omega_R \leq \Delta_L + \Delta_S$ and the peak at $\omega_R$ is the spin resonance, (2) $\omega_R > 2\Delta_L$ and the peak is definitely not a spin resonance, and (3) $\omega_R \leq 2\Delta_L$ and the peak is most likely the spin resonance but the definitive conclusion can be drawn only from the calculation of the dynamical spin susceptibility for the particular experimental band structure.

\begin{acknowledgements}
We would like to thank H. Kontani, S.A. Kuzmichev, T.E. Kuzmicheva, V.M. Pudalov, and I.S. Sandalov for useful discussions. MMK is grateful to B. Keimer and Max-Planck-Institut f\"{u}r Festk\"{o}rperforschung for the hospitality during his visit. We acknowledge partial support by RFBR (grant 16-02-00098), and Government Support of the Leading Scientific Schools of the Russian Federation (NSh-7559.2016.2).
\end{acknowledgements}

\clearpage

\section{Supplemental material for the article ``Spin resonance peak in Fe-based superconductors with unequal gaps'' \label{suppl}}

\textit{Review of experimental data on the peak in inelastic neutron scattering (INS) and gaps extracted from various experimental techniques.}

In the article, we have analysed the spin response of FeBS with two different superconducting gap scales, $\Delta_L > \Delta_S$. Spin resonance appears in the $s_\pm$ state below the indirect gap scale $\tilde\Delta$ that is determined by the sum of gaps on two different Fermi surface sheets connected by the scattering wave vector $\Q$. For the Fermi surface geometry characteristic to the most of FeBS materials, the indirect gap is either $\tilde\Delta = \Delta_L + \Delta_S$ or $\tilde\Delta = 2\Delta_L$. This gives the simple criterion to determine whether the experimentally observed peak in inelastic neutron scattering is the true spin resonance -- if the peak frequency $\omega_R$ is less than the indirect gap $\tilde\Delta$, then it is the spin resonance and, consequently, the superconducting state has the $s_\pm$ gap structure.

Sometimes it is not always clear experimentally which gaps are connected by the wave vector $\Q$. Even without knowing this exactly, one can draw some conclusions. For example, if one of the gaps is $\Delta_L$, then there are three cases possible: (1) $\omega_R \leq \Delta_L + \Delta_S$ and the peak at $\omega_R$ is the spin resonance, (2) $\omega_R > 2\Delta_L$ and the peak is definitely not a spin resonance, and (3) $\omega_R \leq 2\Delta_L$ and the peak is most likely the spin resonance but the definitive conclusion can be drawn only from the calculation of the dynamical spin susceptibility for the particular experimental band structure.

Here we combine data on the peak frequency $\omega_R$ and maximal and minimal gap sizes $\Delta_L$ and $\Delta_S$ available in the literature. Results are presented in Table~\ref{tab}. Unfortunately, for many materials either the INS data or gaps estimations are absent. This gives a whole set of tasks for future experiments. Here are some conclusions, which we can make:
\begin{enumerate}
  \item In electron-doped BaFe$_{1-x}$Co$_{x}$As$_2$ system, NaFe$_{1-x}$Co$_{x}$As system, and FeSe, $\omega_R < \Delta_L + \Delta_S$ and, thus the peak in INS is the true spin resonance evidencing sign-changing gap.

  \item Some hole doped Ba$_{1-x}$K$_{x}$Fe$_2$As$_2$ materials satisfy $\omega_R \leq \Delta_L + \Delta_S$ condition, and some satisfy $\omega_R < 2\Delta_L$ condition. Latter comes especially from newer tunneling~\cite{_Shan2012,_Shimojima2011} and Andreev reflection~\cite{_Abdel-Hafiez2014} data reveling smaller gap values. The fact that $\omega_R < 2\Delta_L$ is still consistent with the sign-changing gap, but as we mentioned before, the calculation of the spin response for the particular experimental band structure is required to make a final conclusion.

  \item The only case when $\omega_{INS}>2\Delta_L$ is FeTe$_{0.5}$Se$_{0.5}$. According to our analysis, there should be no sign-changing gap structure. But before concluding this since this is the single case only, gap data coming from $\mu$SR~\cite{_Biswas.PhysRevB.81.092510.pdf,_Bendele.PhysRevB.81.224520.pdf} should be double checked by independent techniques.

  \item Interesting to note, that ARPES in all cases gives gaps values larger than extracted from other techniques. Natural question arise -- whether the ARPES overestimates or all other methods underestimates superconducting gaps?
\end{enumerate}

\begin{table*}
\caption{\label{tab} Comparison of peak energies in INS and larger and smaller gaps in various FeBS. Values of $\omega_{INS}$, $\Delta_L$, and $\Delta_S$ are given in meV, $*$ marks the Andreev reflections data, $\dag$ marks gaps extracted from tunneling spectra and STS, $\ddag$ marks gaps extracted from optical spectroscopy, $\dag\dag$ marks gaps extracted from muon spin rotation ($\mu$SR), $**$ marks gaps extracted from the BCS fit of $H_{c1}(T)$, otherwise gaps are extracted from ARPES. If the peak frequency and gaps satisfy condition \DLS{$\omega_{INS}<\Delta_L+\Delta_S$}, gaps are marked by \DLS{green} color, and if they satisfy condition \DLL{$\omega_{INS}<2\Delta_L$}, gaps are marked by \DLL{yellow} color. \GDLL{Red} color is used in the case of \GDLL{$\omega_{INS}>2\Delta_L$}.}
\begin{tabular}{cccccc}
\hline
\hline
\centering{Material} & $T_c$ (K) & $\omega_{INS}$ & $\Delta_L$, $\Delta_S$ \\
\hline
BaFe$_{1.9}$Co$_{0.1}$As$_2$ & 19 & 8.3~\cite{_Wang2016} & \DLS{5.0, 4.0}~\cite{_Wang2016} \\
BaFe$_{1.866}$Co$_{0.134}$As$_2$ & 25 & 8.0~\cite{_Wang2016} & \DLS{6.5, 4.6}~\cite{_Wang2016} \\
BaFe$_{1.81}$Co$_{0.19}$As$_2$ & 19 & 8.5~\cite{_Wang2016} & \DLS{5.6, 4.6}~\cite{_Wang2016} \\
BaFe$_{1.85}$Co$_{0.15}$As$_2$ & 25 & 9.5~\cite{_Inosov2010,_Park.PhysRevB.82.134503.pdf} & \DLS{6.7, 4.5}~\cite{_Terashima2009} \\
BaFe$_{1.85}$Co$_{0.15}$As$_2$ & 25.5 & $\sim 9.5$? & \DLS{6.6, 5}~\cite{_Kawahara2010} \\
BaFe$_{1.8}$Co$_{0.2}$As$_2$ & 24.5 & $\sim 9.5$? & \DLS{9, 4}*~\cite{_Tortello2010} \\
BaFe$_{1.85}$Co$_{0.15}$As$_2$ & 25.3 & ? & 5.52-6.98$^\dag$~\cite{_Yin2009} \\
BaFe$_{1.84}$Co$_{0.16}$As$_2$ & 22 & 8.6~\cite{_Lumsden2009} & 7$^\dag$~\cite{_Massee2009} \\
BaFe$_{1.9}$Ni$_{0.1}$As$_2$ & 20 & 9.1~\cite{_Chi2009} & ? \\
BaFe$_{1.91}$Ni$_{0.09}$As$_2$ & 18 & 6.5-8.7~\cite{_Park.PhysRevB.82.134503.pdf} & ? \\
Ba(Fe$_{0.65}$Ru$_{0.35}$)$_2$As$_2$ & 20 & 8~\cite{_Zhao.PhysRevLett.110.147003.pdf} & ? \\
Ba$_{0.6}$K$_{0.4}$Fe$_2$As$_2$ & 38 & 14~\cite{_ChristiansonBKFA,_Castellan2011,_Shan2012} & \DLS{12.5, 5.5}~\cite{_Ding2008,_Wray2008} \\
Ba$_{0.6}$K$_{0.4}$Fe$_2$As$_2$ & 38 & 14~\cite{_ChristiansonBKFA,_Castellan2011,_Shan2012} & \DLS{7-11.5, 4-7}~\cite{_Zhang2010} \\
Ba$_{0.6}$K$_{0.4}$Fe$_2$As$_2$ & 38 & 14~\cite{_ChristiansonBKFA,_Castellan2011,_Shan2012} & \DLL{8.4, 3.2}$^\dag$~\cite{_Shan2012,_Shimojima2011} \\
Ba$_{0.6}$K$_{0.4}$Fe$_2$As$_2$ & 35 & 14~\cite{_Castellan2011} & \DLS{10-12, 7-8}~\cite{_Zhao2008} \\
Ba$_{0.6}$K$_{0.4}$Fe$_2$As$_2$ & 37.5 & 14~\cite{_Castellan2011} & \DLL{8.5-9.3, 1.7-2.3}**~\cite{_Ren2008} \\
Ba$_{0.67}$K$_{0.33}$Fe$_2$As$_2$ & 38 & 15~\cite{_Zhang2011} & ? \\
Ba$_{0.65}$K$_{0.35}$Fe$_2$As$_2$ & 34 & 13~\cite{_Castellan2011} & \DLL{7.4-8, 1.4-2}*~\cite{_Abdel-Hafiez2014} \\
Ba$_{1-x}$K$_{x}$Fe$_2$As$_2$ & 32 & 14~\cite{_Castellan2011} & \DLL{9.2, 1.1}~\cite{_Evtushinsky2009,_Evtushinsky2009NJP} \\
Ba$_{0.55}$K$_{0.45}$Fe$_2$As$_2$ & 23 & ? & 9.2, 2.7*~\cite{_Samuely2009} \\
Ba$_{0.3}$K$_{0.7}$Fe$_2$As$_2$ & 22 & ? & 7.9, 4.4~\cite{_Nakayama.PhysRevB.83.020501.pdf} \\
Ba$_{0.1}$K$_{0.9}$Fe$_2$As$_2$ & 9 & ? & 2.7-3.6~\cite{_Xu.PhysRevB.88.220508.pdf} \\
K$_{0.8}$Fe$_2$Se$_2$ & 31 & ? & 10.3~\cite{_Zhang.nmat2981.pdf} \\
Cs$_{0.8}$Fe$_2$Se$_2$ & 30 & ? & 10.3~\cite{_Zhang.nmat2981.pdf} \\
FeSe & 8 & 4~\cite{_Wang.nmat4492.pdf} & \DLS{2.5, 3.5}$^\dag$~\cite{_Kasahara.PNAS.111.16309.pdf} \\
FeSe & 8 & 4~\cite{_Wang.nmat4492.pdf} & \DLS{0.6-1, 2.4-3.2}*~\cite{_Ponomarev2013} \\
FeTe$_{0.5}$Se$_{0.5}$ & 14 & 6-7~\cite{_Mook.PhysRevLett.104.187002.pdf,_Lee2010,_Wen2010} & \GDLL{2.61, 0.51-0.87}$^{\dag\dag}$~\cite{_Biswas.PhysRevB.81.092510.pdf,_Bendele.PhysRevB.81.224520.pdf} \\
FeTe$_{0.55}$Se$_{0.45}$ & 14 & ? & 5.1, 2.5$^\ddag$~\cite{_Homes.PhysRevB.81.180508.pdf,_Homes.JPhysChemSol.72.505.pdf} \\
FeTe$_{0.6}$Se$_{0.4}$ & 14 & 6.5~\cite{_Qiu2009,_Argyriou2010} & ? \\
Fe$_{1.03}$Te$_{0.7}$Se$_{0.3}$ & 13 & ? & 4~\cite{_Nakayama.PhysRevLett.105.197001.pdf} \\
%
%
BaFe$_2$(As$_{0.65}$P$_{0.35}$)$_2$ & 30 & 12~\cite{_Ishikado2011,_Ishikado2011PRB} & ? \\
LaFeAsO$_{0.92}$F$_{0.08}$ & 29 & 13~\cite{_Shamoto2010} & ? \\
LaFeAsO$_{0.943}$F$_{0.057}$ & 25 & 11-12~\cite{_Wakimoto2010} & ? \\
LaFeAsO$_{1-x}$F$_{x}$ & 22 & ? & 5.4, 1.4*~\cite{_Kuzmichev2016} \\
LiFeAs & 18 & 8~\cite{_Taylor.PhysRevB.83.220514.pdf} & \DLS{5-6, 2.8-3.5}~\cite{_Borisenko.PhysRevLett.105.067002.pdf,_Borisenko.symmetry-04-00251.pdf,_Umezawa.PhysRevLett.108.037002.pdf} \\
LiFeAs & 18 & 8~\cite{_Taylor.PhysRevB.83.220514.pdf} & \DLL{5.4, 1.4}*~\cite{_Kuzmichev2012_,Kuzmichev2013} \\
LiFeAs & 18 & 8~\cite{_Taylor.PhysRevB.83.220514.pdf} & \DLL{5.3, 2.5}$^\dag$~\cite{_Chi.PhysRevLett.109.087002.pdf,_Hanaguri.PhysRevB.85.214505.pdf,_Nag.srep27926.pdf} \\
%
NaFe$_{0.978}$Co$_{0.022}$As & 18 & 7.5~\cite{_Charnukha.SciRep.5.18620.(2015).pdf} & 2.8$^\ddag$~\cite{_Charnukha.SciRep.5.18620.(2015).pdf} \\
NaFe$_{0.935}$Co$_{0.045}$As & 18 & 7~\cite{_Zhang.PhysRevB.88.064504.pdf} & \DLS{6.8, 6.5}~\cite{_Liu.PhysRevB.84.064519.pdf} \\
NaFe$_{0.95}$Co$_{0.05}$As & 18 & $\sim 7$? & \DLS{6.8, 6.5}~\cite{_Liu.PhysRevB.84.064519.pdf} \\
NdFeAsO$_{0.9}$F$_{0.1}$ & 53 & ? & 15~\cite{_Kondo.PhysRevLett.101.147003.pdf}\\
Tl$_{0.63}$K$_{0.37}$Fe$_{1.78}$Se$_2$ & 29 & ? & 8.5~\cite{_Wang.epl.93.57001.pdf} \\
\hline
\hline
\end{tabular}
\end{table*}

\end{document}